\documentclass[useAMS,usenatbib,usegraphicx]{mn2e}

\usepackage{color}
\usepackage{soul}

\title[A tomographic study of V691 CrA (X1822-371)]{A tomographic study of V691 CrA (X1822-371)$^{}$\thanks{This paper includes data gathered with the 6.5 m Magellan Telescopes located at Las Campanas Observatory, Chile.}}
\author[C. S. Peris and S. D. Vrtilek]{C. S. Peris$^{1,2}$\thanks{E-mail:
cperis@cfa.harvard.edu, c.peris@neu.edu} and S. D.
Vrtilek$^{1}$\thanks{E-mail: svrtilek@cfa.harvard.edu}\\
$^{1}$Harvard-Smithsonian Center for Astrophysics, 60 Garden Street, Cambridge, MA 02138, U.S.A.\\
$^{2}$Department of Physics, Northeastern University, Boston, MA 02115, U.S.A.}
\begin{document}

\date{Accepted ???? Received ???? in original form ????}

\pagerange{\pageref{firstpage}--\pageref{lastpage}} \pubyear{2012}

\maketitle

\label{firstpage}

\begin{abstract}
We present Doppler and modulation tomography of the low-mass X-ray binary V691 CrA with data obtained using the 6.5-m Magellan Baade telescope at the Las Campanas Observatory in 2010 and 2011.  The disc and hotspot are observed in H$\alpha$ and He II ($\lambda$4686) in both years. A clear image of the disc is seen in He II ($\lambda$5411) using the 2010 data. We present the first tomography of the absorption line He I ($\lambda$5876) and detect absorption near the L$_1$ point of the donor star. We also present the first modulation tomography of the emission line H$\alpha$ and detect emission from the secondary. The H$\alpha$ double peaks are imbedded in a deep absorption trough confirming the presence of Balmer line absorbing material in the system. Our observations of H$\alpha$ show absorption in a larger phase range than in H$\beta$ which could be due to heating up of sprayed matter from the hotspot as it travels downstream. We also suggest possible occultation of the H$\alpha$ absorbing spray by the disc bulge at certain phases.
\end{abstract}


\section{Introduction}

V691 Coronae Australis (V691 CrA; X1822-371) is a low-mass X-ray binary (LMXB) with an orbital period of 5.57 hr \citep{b15}. Although it has a high inclination (i = 82.5$\,^{\circ}$$\pm$1.5$\,^{\circ}$; Heinz \& Nowak 2001), it shows only partial eclipses indicating the presence of an accretion disc corona (ADC; White $\&$ Holt 1982). The detection of 0.59 s pulsations in X-rays allowed for an accurate determination of the neutron star radial velocity semi-amplitude, K$_1$ = 94.5 km s$^{-1}$ \citep{b17}. Estimates of the masses of the neutron star and the secondary range from 1.14 M$_{\sun}$ $\leq$ M$_1$ $\leq$ 2.32 M$_{\sun}$ and 0.36 M$_{\sun}$ $\leq$ M$_2$ $\leq$ 0.56 M$_{\sun}$ \citep{b6,b22}.

The short orbital period and optical brightness of V691 CrA, although compromised by the eclipses, make it a good candidate for tomography \citep{b18,b19,b28}. \citet{b6} produced Doppler maps in He II ($\lambda$4686), O VI ($\lambda$3811), He I ($\lambda$4471) and the N III ($\lambda$4640) line of the Bowen blend. They found that the N III emission was associated with the secondary whereas He II arose from the disc. \citet{b30} using tomograms in He II ($\lambda$4686 and $\lambda$5411) and the Bowen blend suggest that the line emission is more likely to come from above the accretion disc, possibly from a disc wind.

We observed V691 CrA in 2010 and 2011 8-9 years after \citet{b6} and 4-5 years after \citet{b30} and generated tomograms in all of the above mentioned lines. Our observations covered a greater wavelength range ($\sim$ $\lambda$4400-$\lambda$6650) than previous tomographic studies allowing us to present the first modulation tomograms of the emission line H$\alpha$ and the absorption line He I ($\lambda$5876). In section 2 we describe the optical observations obtained and used for this study. In section 3 we present our analysis of the spectral features. In section 4 we present Doppler and modulation tomography of the lines and discuss the conclusions drawn from the analysis in section 5.

\section[]{Observations}

We observed V691 CrA with IMACS on the 6.5 m Magellan Baade telescope at the Las Campanas Observatory. We used a 600 lines/mm grating which gave a velocity dispersion of 37 km s$^{-1}$. 44 spectra were obtained on 2010 June 7-8 and 24 spectra on 2011 June 28 with total integration times of 4.6167 hr and 4 hr respectively and exposure times as given in Table 1. Atmospheric conditions required slits of 0.75" and 1" for observations made in 2010 while all observations made in 2011 were done with a slit width of 0.9" with corresponding spectral resolutions of 2.55, 3.4 and 3.06 \AA.

\begin{table}
\caption{Observation log}
\label{log}
\begin{tabular}{@{}lccc}
\hline
Date & $\#$ of spectra
& Exposure(s)\\
\hline
2010 June 7-8 & 44 & 360/420\\
2011 June 28 & 24 & 600\\
\hline
\end{tabular}
\end{table}

The spectra were reduced using standard image reduction routines in IRAF version 2.10 (zero correction, flat fielding and spectral extraction). The onedspec format was used to extract the spectra. After cleaning and wavelength calibration, the spectra were converted into ASCII format to be read into the software package MOLLY, a 1D spectrum analysis program developed by Tom Marsh.

\section[]{Analysis}

Using MOLLY the spectra were corrected for the heliocentric effects, and normalized by fitting polynomials of the order 10 to the continuum and subtracting these polynomials from the corresponding spectra.
We used the quadratic ephemeris given by \citet{b36} to calculate the phase of each spectrum. We used the most recent ephemeris for the eclipse given by \citet{b3} using Chandra data taken in 2008 as our value for T$_0$. The gap between the observation date and this ephemeris consisted of 3223 cycles for the 2010 data and 4881 cycles for the 2011 data and gave us phase errors of 0.0032 cycles and 0.0037 cycles respectively.

The ephemeris was defined using:

\[
\begin{array}[b]{r}
T = 2452607.70052(56) + 0.232108785(50)\bmath{E} \\
     + 2.06(23)\times10^{-11}\bmath{E}^{2}
\end{array}
\]

\begin{figure*}
%\begin{minipage}{130}
\includegraphics[width=130mm, trim = 10 130 10 130, clip = true]{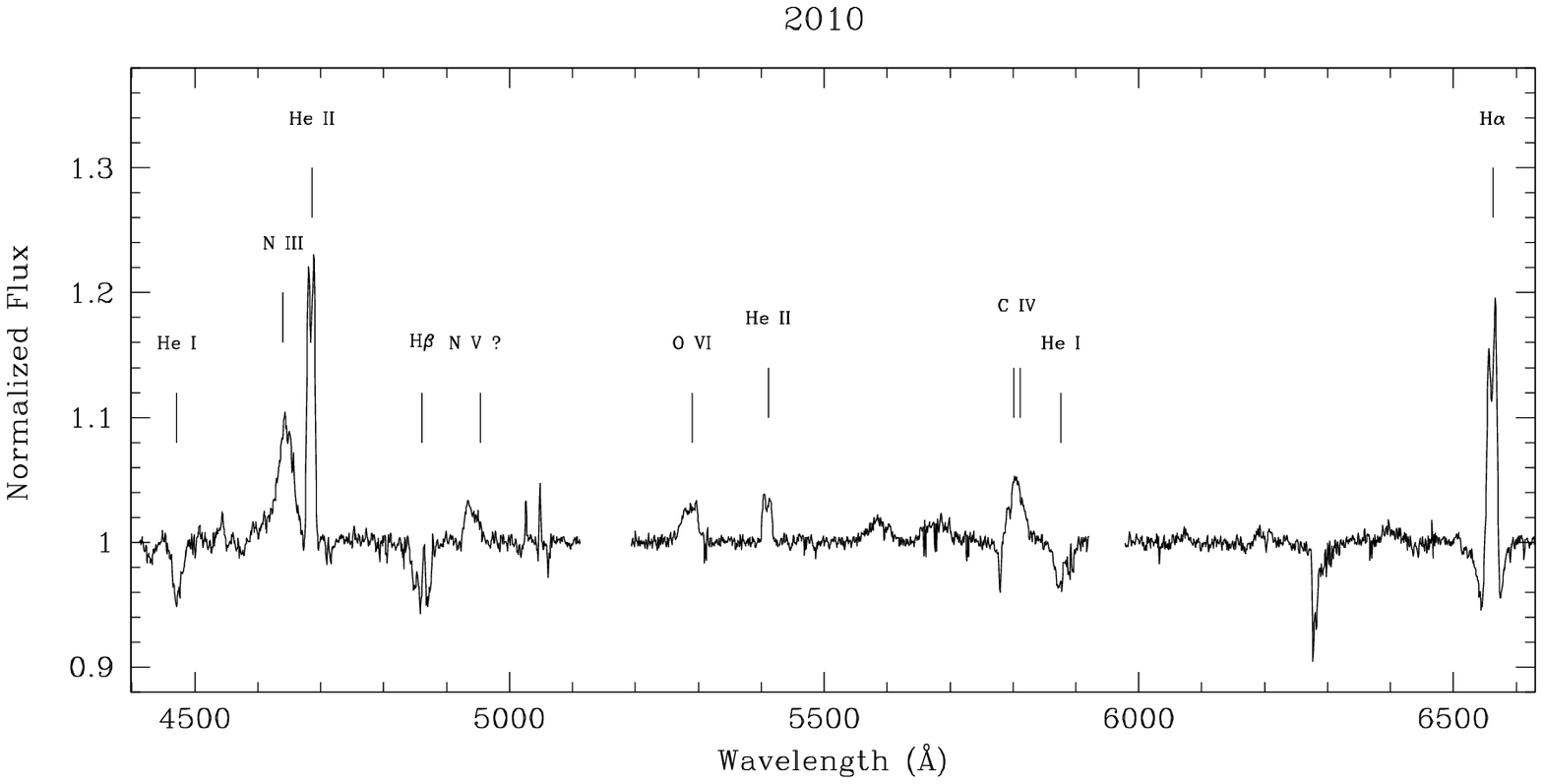}
\includegraphics[width=130mm, trim = 10 130 10 130, clip = true]{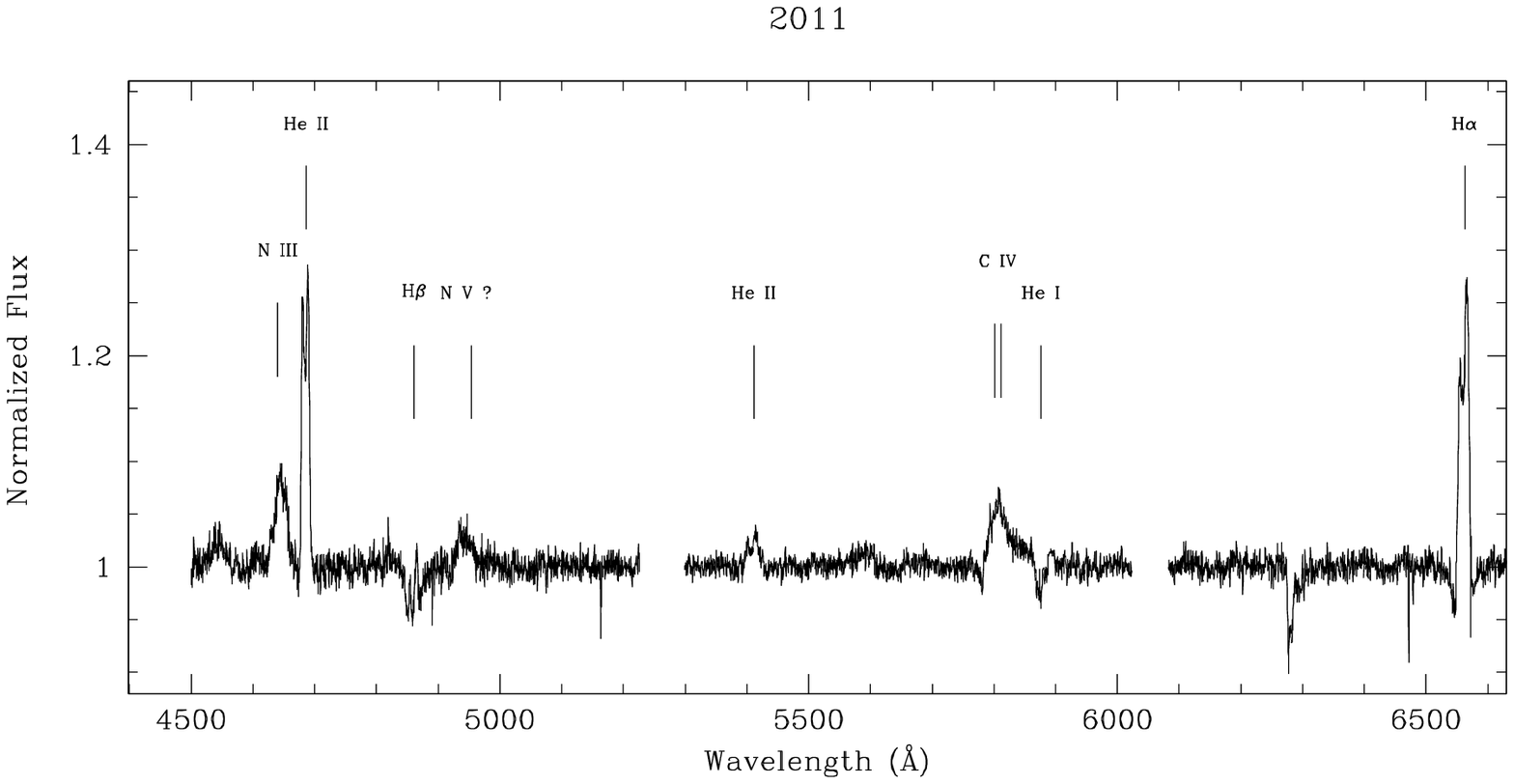}
%\vspace{250pts}
\caption{The normalized average spectra of the system using 2010 and 2011 data.
Main spectral features are indicated. The feature at $\lambda$6278 is
contaminated by telluric O$_2$ lines.}
\label{}
%\end{minipage}
\end{figure*}

\subsection[]{Spectral features and profiles}

\begin{figure*}
%\begin{minipage}{170}
\includegraphics[width=85mm]{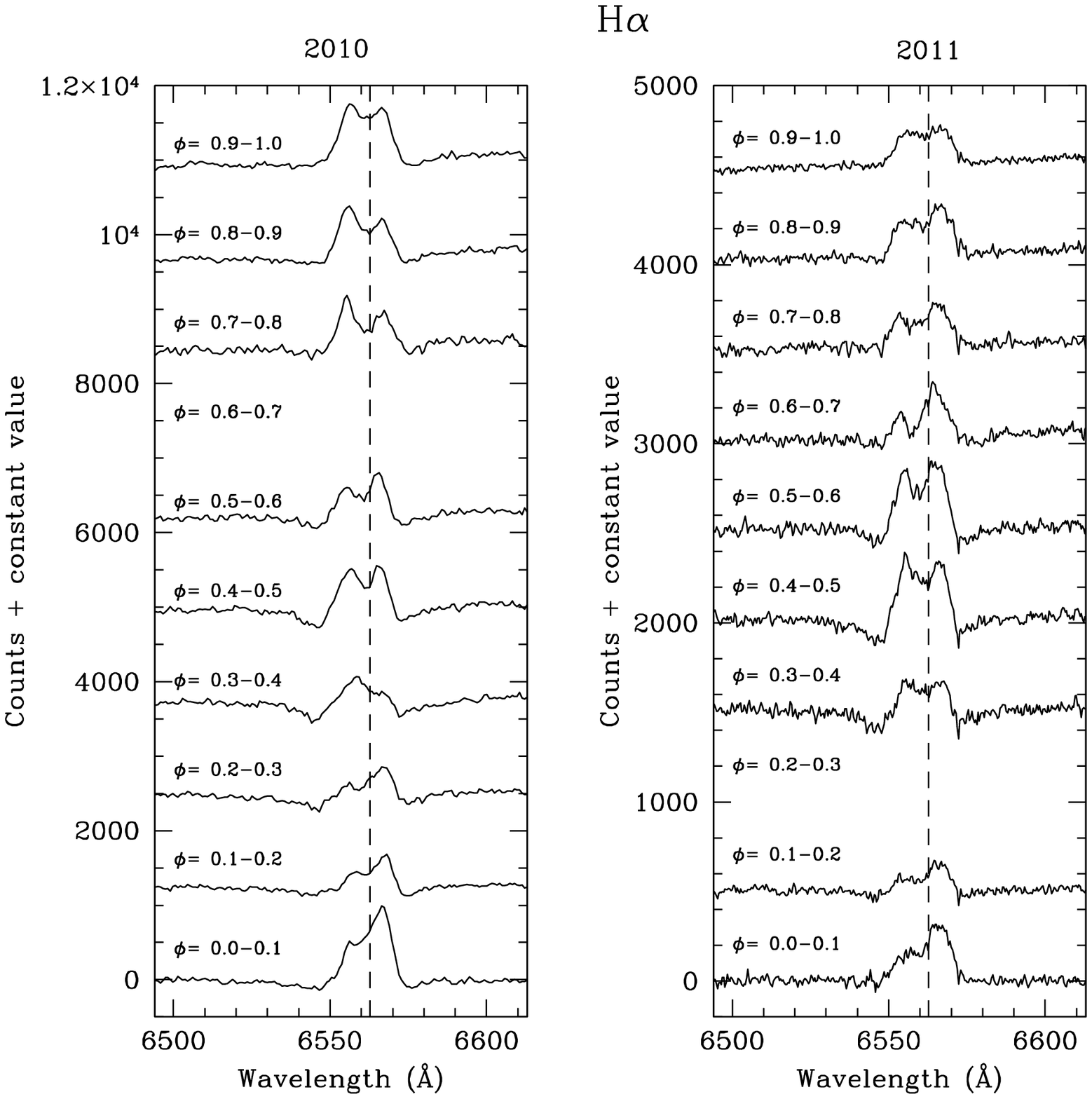}
\includegraphics[width=85mm]{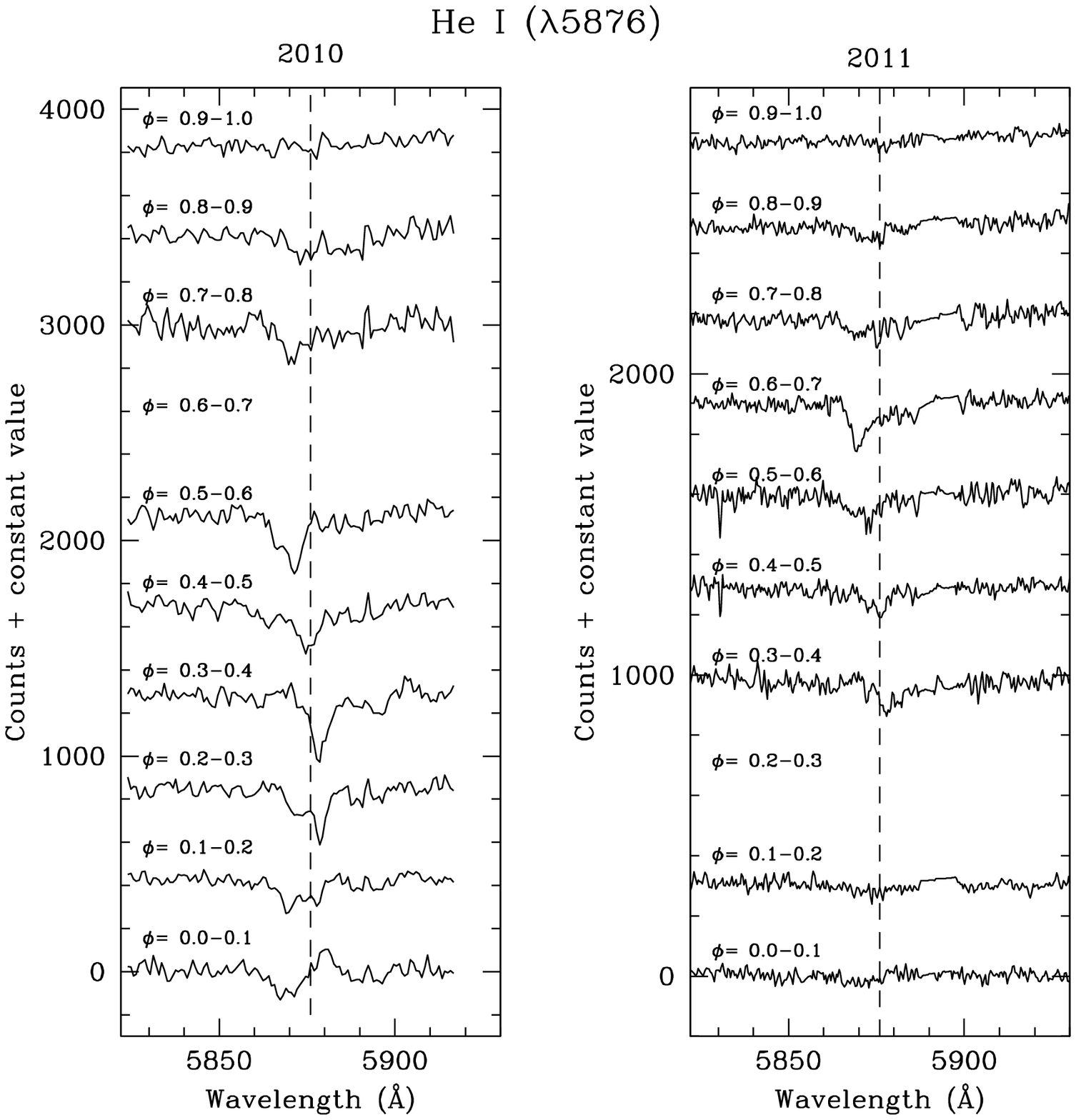}
\includegraphics[width=85mm]{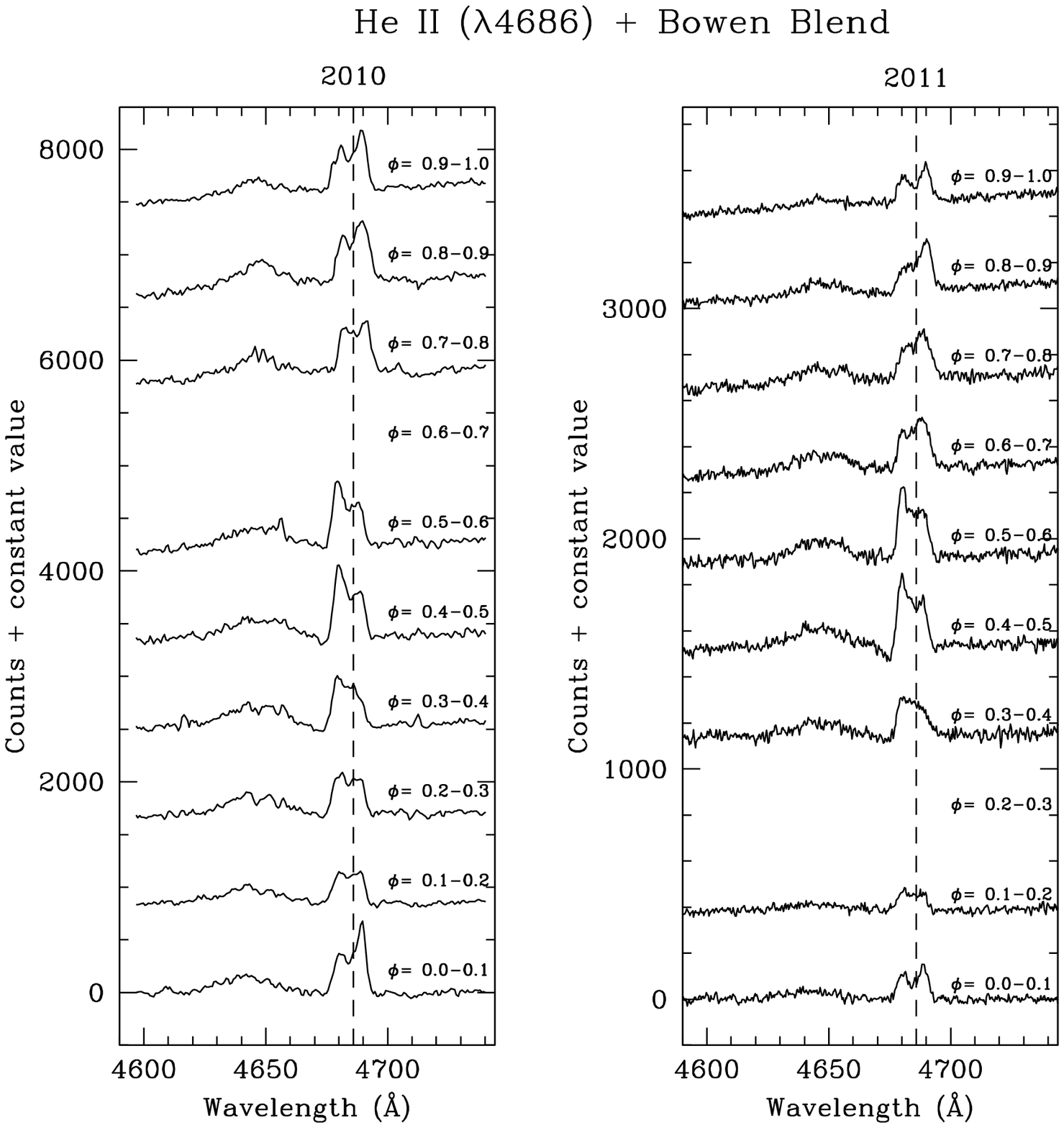}
\includegraphics[width=85mm]{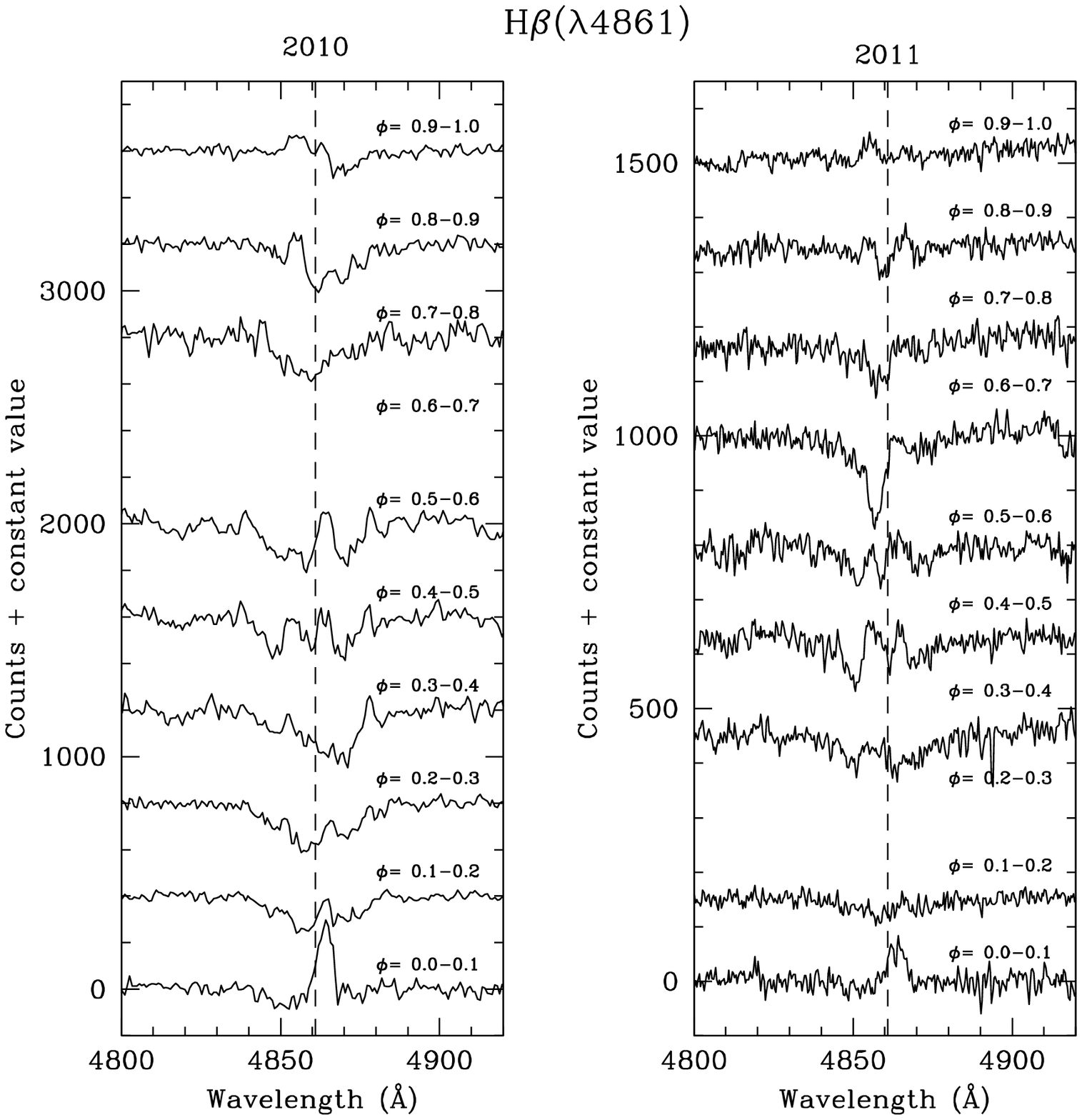}
%\vspace{250pts}
\caption{H$\alpha$, He II (4686), He I (5876) and H$\beta$ spectra in 10 phase bins showing the evolution of the lines over the orbital period.}
\label{}
%\end{minipage}
\end{figure*}

The average spectra (Fig. 1) show strong emission in H$\alpha$ and He II ($\lambda$4686) with double peak profiles typically seen in lines originating from the disc. They also display the absorption line He I ($\lambda$5876) which trailed data showed as having a sinusoidal variation. The 2010 data displays a double peak profile in He II ($\lambda$5411), a prominent Bowen blend and an emission feature at $\lambda$5290 which was suggested to be O VI by \citet{b5} and \citet{b30}. Also prominent in the 2010 average spectrum are deep absorption troughs around H$\alpha$ and H$\beta$. These absorption troughs are observed to decrease in depth from 2010 to 2011. 

Fig. 2 shows the evolution of the main spectral features over the orbital period. He II ($\lambda$4686) displays the classic double peak variation as would be expected from emission from an accretion disc while the Bowen blend exhibits some binary motion as well. The He I ($\lambda$5876) peak profile is observed to be relatively dim at phases 0.8-0.9 and 0.9-1.0 and somewhat complex at phase 0.0-0.1.

\subsection{Complex profile in Balmer lines}
   
H$\alpha$ is the strongest feature in our spectra but it has a complex profile with the emission imbedded in an absorption trough. The absorption can be observed in all phases while peak absorption occurs at phases 0.2 - 0.6 (Fig. 3). This is consistent with the behavior of H$\alpha$ observed by \citet{b13}. We also observe a complex profile in H$\beta$ which has been previously noted by \citet{b30} and \citet{b6}. The evolution of H$\beta$ over the orbital period is similar to that observed by \citet{b30} except for a blueshifted absorption feature at phases 0.2 - 0.6.

\section{Tomography}
\subsection{Doppler Tomography}

Doppler tomography uses the information encoded in the line profiles as a function of the orbital phase to calculate the strength of emission as a function of velocity. The Doppler tomograms we present were created using the maximum entropy (MEM) method. A grid of pixels spanning velocity space is adjusted to achieve a target goodness-of-fit measured by $\chi$$^{2}$, while simultaneously maximizing the entropy of the image. A refined form of entropy which measures departures from a 'default' image is used. For an excellent summary of the method, see \citet{b37}.

For the Doppler tomography of V691 CrA we used the systemic velocity of $\gamma$ = -43 km s$^{-1}$ computed by \citet{b6} which was verified using the optgam feature in the Doppler software. Doppler tomograms of H$\alpha$ ($\lambda$6562.76), He II ($\lambda$4686), He I ($\lambda$5876) are shown in Figs 4-9.

The Doppler tomogram in H$\alpha$ produced using the 2010 data (Fig. 3) shows a somewhat elongated disc with the emission concentrated in the positive V$_y$ region. Two bright spots are observed: one on the positive V$_y$ axis, the other on the positive V$_y$ and negative V$_x$ quadrant. The positioning of the latter suggests that it originates from the hotspot formed due to the large amount of radiation emitted at the impact site of the accretion stream on the disc \citep{b29}.

Fig. 4 which shows the Doppler tomogram in H$\alpha$ produced using the 2011 data, displays two bright regions: one on the positive V$_y$ axis, the other on the negative V$_y$ and negative V$_x$ quadrant. The bright spots on the positive V$_y$ axis in both Fig. 3 and Fig. 4 lie on the disc but also coincide with the usually accepted position of the secondary.

The Doppler tomogram in He II ($\lambda$4686) created using the 2010 data (Fig. 5) displays a ring structure which appears to be circular with a crescent shaped bright spot that runs along the negative V$_x$ side of the disc. The hotspot can be observed at the upper end of the crescent shaped bright spot. It was observed to lie in a similar region in velocity space to the observed hotspot in the H$\alpha$ Doppler map created using the 2010 data.

The tomogram in He II ($\lambda$4686) created using the 2011 data also displays a ring structure with a crescent shaped bright spot but the spot was positioned in the negative V$_x$ and negative V$_y$ quadrant (Fig. 6). The crescent shaped spots observed in He II ($\lambda$4686) in the 2010 data and the 2011 data seem to be in the same region in velocity space as the bright spots observed in H$\alpha$ in the same respective years suggesting that He II and H$\alpha$ emission originates from the same processes.

The He I absorption feature detected by \citet{b13} was observed in our spectra and we used this feature to produce Doppler tomograms, removing the Na I interstellar lines and using the cleaned spectra for tomography. In the 2010 data we observe a bright spot with its maximum intensity at approximately V$_y$ = 189 km s$^{-1}$ (Fig. 7). \citet{b6} identified a similar spot in He I ($\lambda$4471) as originating from the accretion stream as it was shifted by a phase of -0.05 from the bright spot produced by the N III line in the Bowen Blend which they interpreted as the irradiated face of the secondary. The bright spot we observe using He I ($\lambda$5876) has a larger spread in velocity with respect to the previous observation and seems to overlap with the predicted position of the L$_1$ point of the secondary.

The He I ($\lambda$5876) Doppler tomogram produced using the 2011 data (Fig. 8) shows a significant shift in the He I absorption feature toward higher V$_y$ velocities, displaying a bright spot with its maximum intensity at approximately V$_y$ = 279 km s$^{-1}$ and V$_x$ = -117 km s$^{-1}$.

\subsection{Modulation Tomography}

Modulation tomography \citep{b28} relaxes the assumption used in Doppler tomography which is that the line flux remains constant and enables the mapping of the line flux variations as a function of orbital period. We used modulation tomography to study V691 CrA using the same systemic velocity of -43 km s$^{-1}$ which was used for Doppler tomography. In over plotting the Roche lobe of the secondary and the position of the primary we used K$_1$ = 94.5 km s$^{-1}$ \citep{b16} and K$_2$ = 324 km s$^{-1}$ \citep{b6} which gave us a mass ratio of q = 0.29. We also plotted the gas stream trajectory and the Keplerian velocity of the disc along the gas stream path.

Upon close examination, the trailed spectra in H$\alpha$ in 2010 (Fig. 9) were found to consist of two superimposed S-curves which seem to be opposite in phase. The first component is similar to a curve expected from the disc. The second out-of-phase component is much thinner. This out-of-phase S-curve component together with the bright spots observed on the V$_y$ axis suggest H$\alpha$ emission from the secondary. The trailed spectra in H$\alpha$ in 2011 (Fig. 10) display similar superimposed S-curves which are dimmer due to the lower number of spectra obtained in 2011. The amplitude variation map in 2010 shows modulation arising from this spot which can be attributed to the sinusoidal variation of the line flux originating from the secondary.

\begin{figure}
\includegraphics[width=84mm,trim = 390 70 380 29, clip = true]{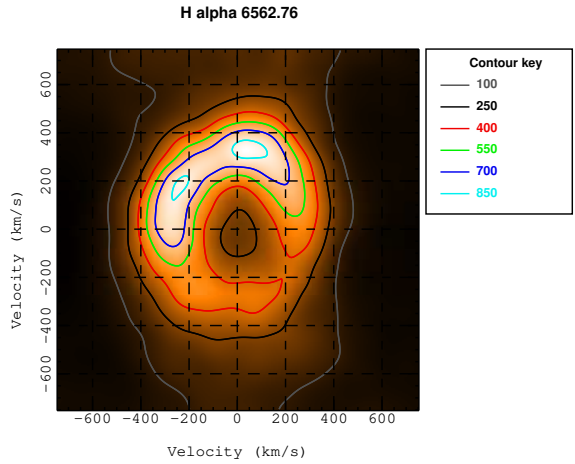}
\caption{Doppler map of H$\alpha$ created using 2010 data}
\label{H_dop10}
\end{figure}

\begin{figure}
\includegraphics[width=84mm,trim = 390 70 380 29, clip = true]{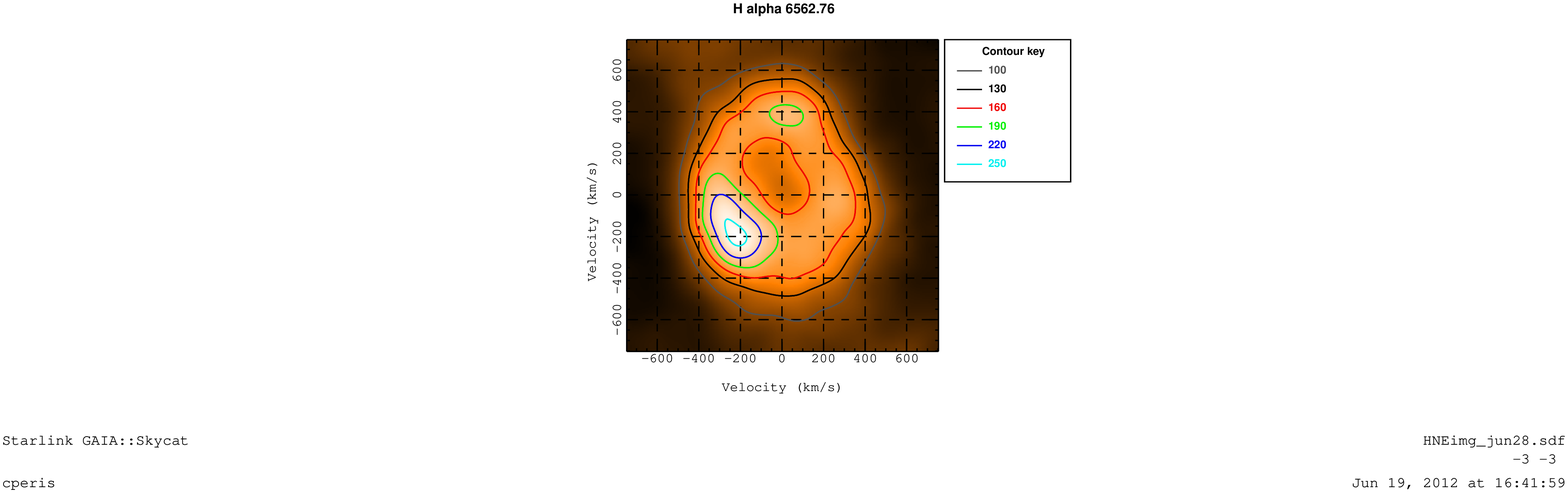}
\caption{Doppler map of H$\alpha$ created using 2011 data}
\label{H_dop11}
\end{figure}

\begin{figure}
\includegraphics[width=84mm,trim = 390 70 380 29, clip = true]{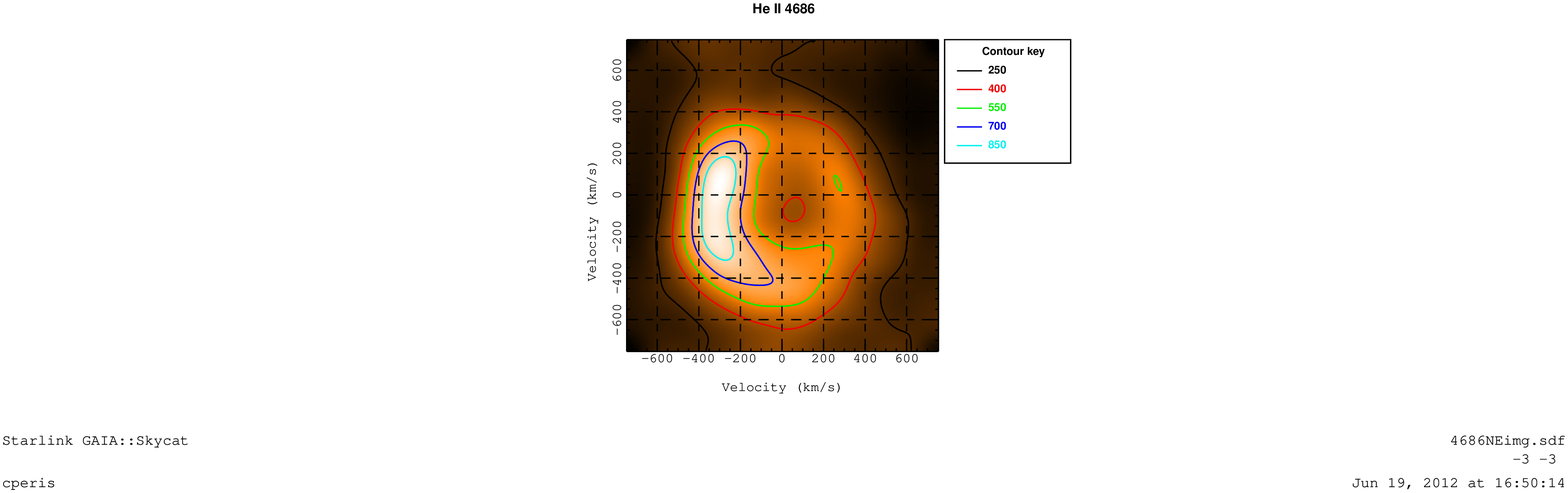}
\caption{Doppler map of He II ($\lambda$4686) created using 2010 data}
\label{He2_dop10}
\end{figure}

\begin{figure}
\includegraphics[width=84mm,trim = 390 70 380 29, clip = true]{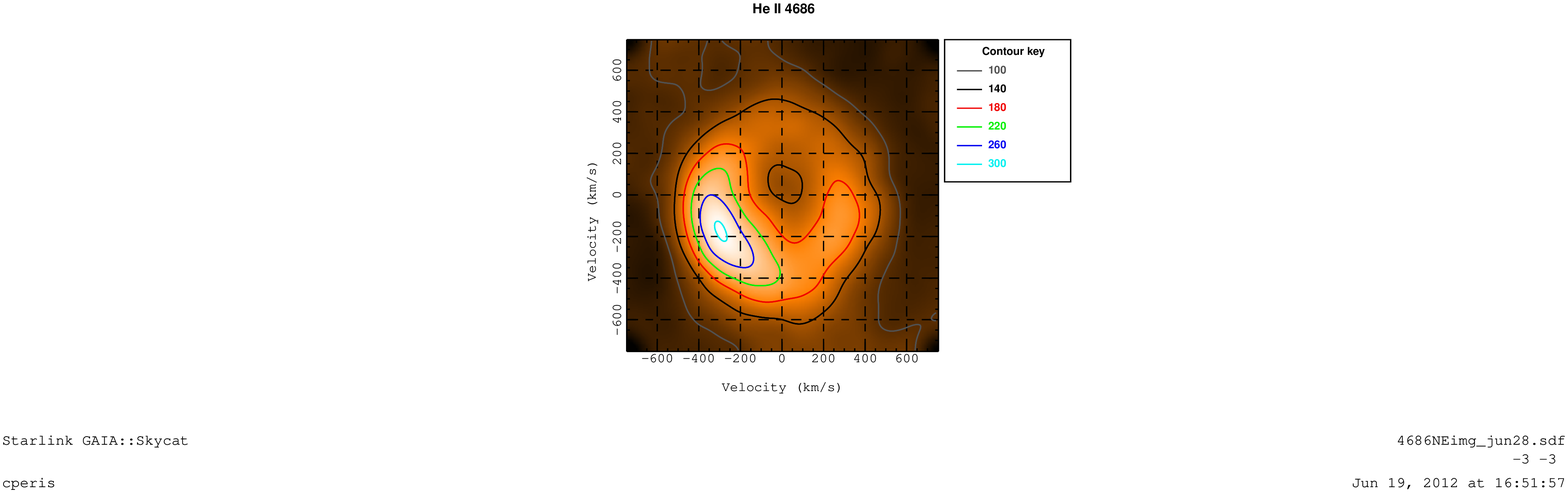}
\caption{Doppler map of He II ($\lambda$4686) created using 2011 data}
\label{He2_dop11}
\end{figure}

\begin{figure}
\includegraphics[width=84mm,trim = 390 70 380 29, clip = true]{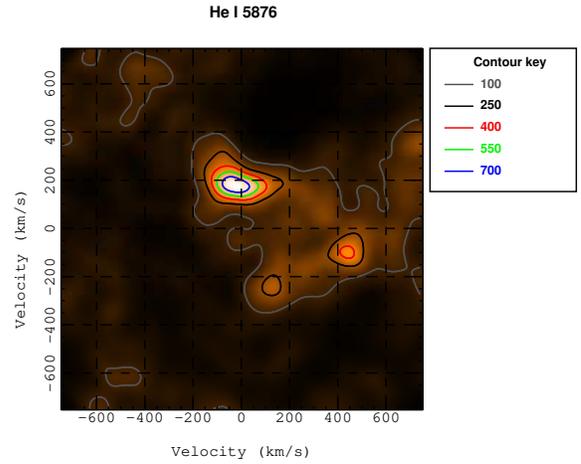}
\caption{Doppler map of He I ($\lambda$5876) created using 2010 data. Since He I ($\lambda$5876) is an absorption line the tomogram was created using inverted spectra.}
\label{He1_dop11}
\end{figure}

\begin{figure}
\includegraphics[width=84mm,trim = 390 70 380 29, clip = true]{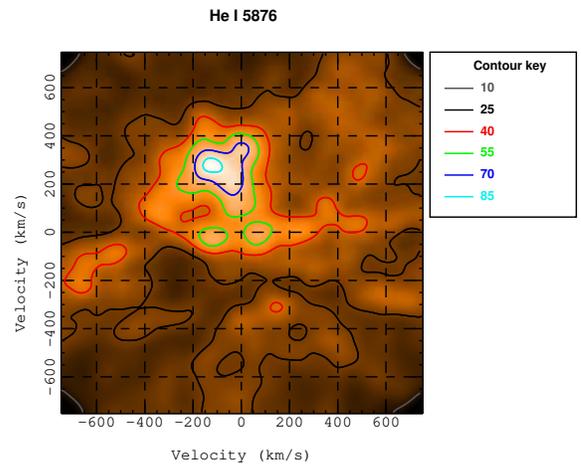}
\caption{Doppler map of He I ($\lambda$5876) created using 2011 data. As before, the tomogram was created using inverted spectra.}
\label{He1_dop11}
\end{figure}

The He II lines were used successfully to produce modulation tomograms (Fig. 11 and Fig. 12) which show the ring structure obtained using Doppler tomography. The tomogram in He II ($\lambda$4686) created using the 2010 data shows modulation at the hotspot. The trailed spectra in both tomograms show distinct broad S-curves typical of lines originating from the entire disc. The S-curve in the 2011 data was observed to have the same phasing as in the 2010 data as expected, but was dimmer due to the lower number of spectra.

The ring structure in He II ($\lambda$5411) (Fig. 13) comprises of two crescent shaped slices of emission, one of which corresponds to the position of the crescent shaped bright spot in $\lambda$4686. The amplitude variation map of $\lambda$5411 shows a bright spot which lies on the gas stream trajectory which is modulation originating from the accretion stream/disc interaction point.

The modulation tomogram created using the absorption line He I ($\lambda$5876) in the 2010 data (Fig. 14) shows the bright spot observed using Doppler tomography on the lower side of the predicted position of the secondary. The S-curve of He I is observed to be in opposite phase to that of He II. This is consistent with the assumption that the He I originates close to the inner face of the secondary while the He II lines originate from the accretion disc around the primary. The tomogram in He I ($\lambda$5876) created using the 2011 data (Fig. 15) displays the He I feature at a higher V$_y$ velocity on the edge of the predicted position of the secondary in the middle of the plotted gas stream trajectory and the Keplerian velocity of the disc along the gas stream path. The S-curve is observed to be similar to that observed in the 2010 data but the absorption is predominantly at phase 0.65. Strong bright spots are seen in the amplitude variation maps of Fig. 14 and Fig. 15. These bright spots indicate sinusoidal variation of the line flux originating from the absorber, usually a feature associated with the donor star.

A feature detected at $\lambda$5290 in the 2010 spectrum was suggested to be O VI by \citet{b5} and \citet{b30}. Tomography on this feature revealed two bright spots, one of which was close to the position of the hotspot (Fig. 16). The hotspot was previously observed by \citet{b6} in O VI ($\lambda$3811). O VI also shows faint traces of the disc emission which can be seen in the trailed spectra.

We generated the modulation tomogram of the Bowen blend (Fig. 17) by centring on the strongest line in N III ($\lambda$4640) and using the entire blend region. The tomogram displays a bright spot in the predicted position of the secondary and is consistent with the results of \citet{b6} and \citet{b30}. The bright spot seems to be positioned toward the centre of the star rather than the L$_1$ point as suggested by \citet{b6}. The ring shaped structure seen by \citet{b30} is observed in our tomogram but the diffuse emission is not.

\begin{figure}
\includegraphics[width=88mm,trim = 35 0 0 0, clip = true, angle=-90]{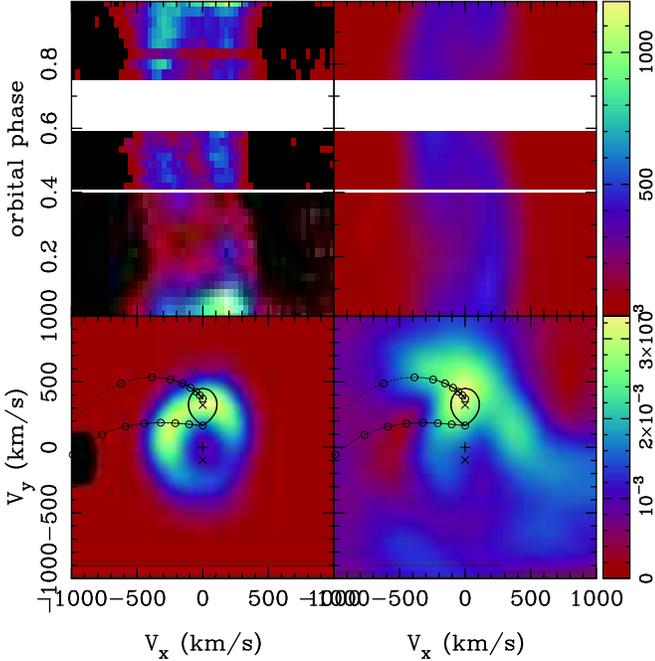}
\caption{Modulation tomogram of H$\alpha$ using 2010 data. The image shows the trailed spectra on the top-left, the simulated trail on the top-right, the system in velocity space on the bottom-left and the amplitude variations on the bottom-right.}
\label{Halpha_2010}
\end{figure}
\begin{figure}
\includegraphics[width=88mm,trim = 35 0 0 0, clip = true, angle=-90]{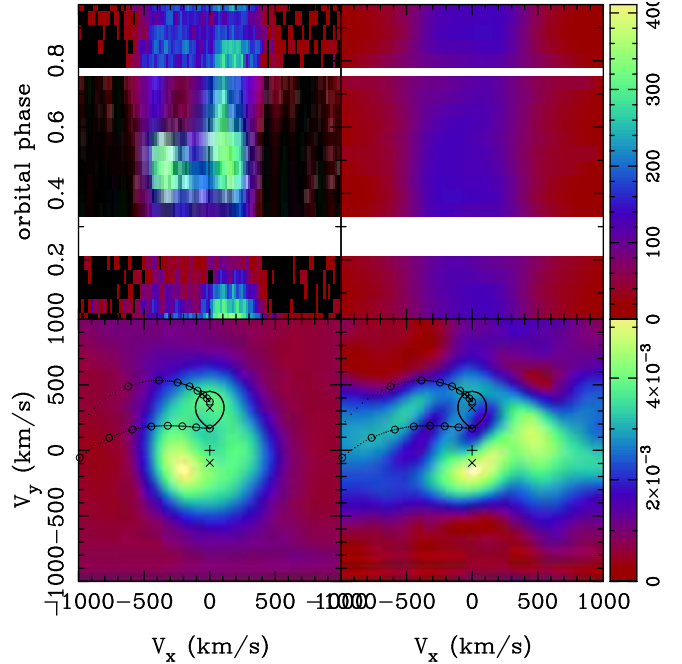}
\caption{Modulation tomogram of H$\alpha$ using 2011 data. Panels are as described in Fig. 9.}
\label{Halpha_2011}
\end{figure}
\begin{figure}
\includegraphics[width=88mm,trim = 35 0 0 0, clip = true, angle=-90]{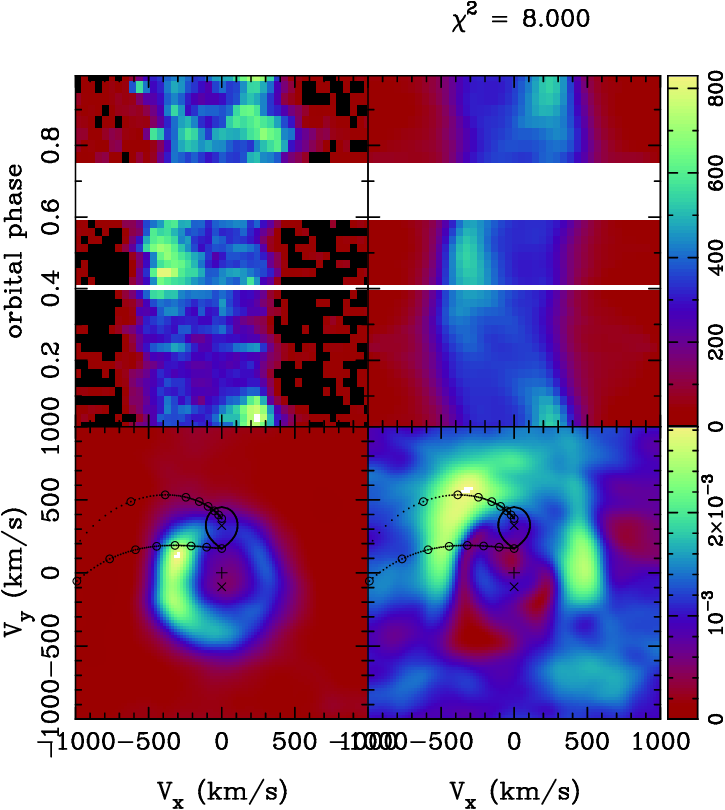}
\caption{Modulation tomogram of He II ($\lambda$4686) in emission using 2010 data. Panels are as described in Fig. 9.}
\label{He2_4686_2010}
\end{figure}
\begin{figure}
\includegraphics[width=88mm, trim = 35 0 0 0, clip = true, angle=-90]{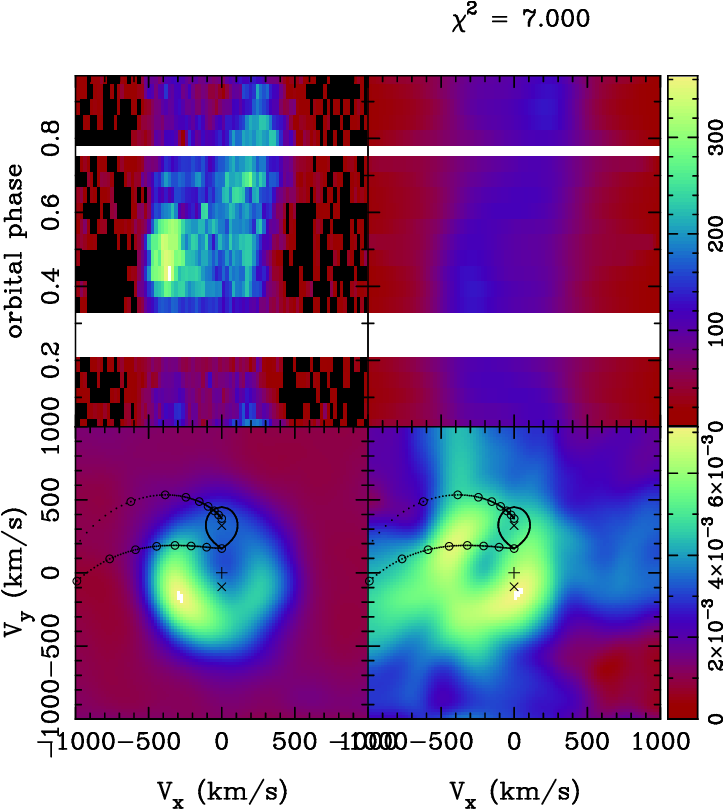}
\caption{Modulation tomogram of He II ($\lambda$4686) in emission using 2011 data. Panels are as described in Fig. 9.}
\label{He2_4686_2011}
\end{figure}
\begin{figure}
\includegraphics[width=88mm, trim = 35 0 0 0, clip = true, angle=-90]{fig13_5411NE2_edit}
\caption{Modulation tomogram of He II ($\lambda$5411) in emission using 2010 data. Panels are as described in Fig. 9.}
\label{He2_5411_2010}
\end{figure}
\begin{figure}
\includegraphics[width=88mm, trim = 35 0 0 0, clip = true, angle=-90]{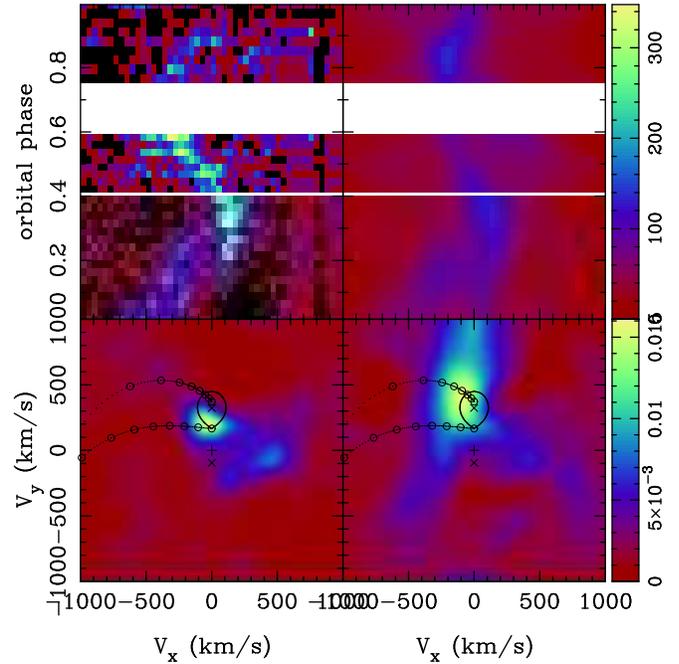}
\caption{Modulation tomogram of the absorption line He I($\lambda$5876) using 2010 data. Panels are as described in Fig. 9. Since He I ($\lambda$5876) is an absorption line the tomogram was created using inverted spectra.}
\label{He1_5876_2010}
\end{figure}
\begin{figure}
\includegraphics[width=88mm, trim = 35 0 0 0, clip = true, angle=-90]{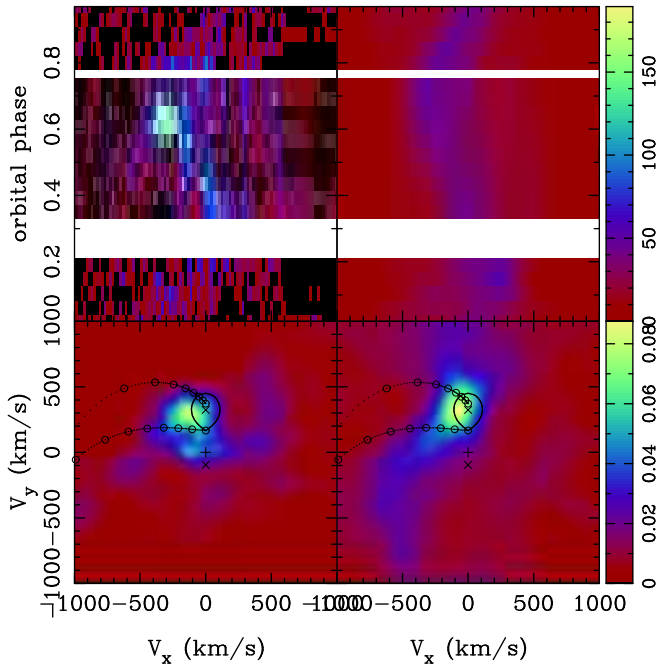}
\caption{Modulation tomogram of the absorption line He I ($\lambda$5876) using 2011 data. Panels are as described in Fig. 9. As in the above case, the tomogram was created using the inverted spectra.}
\label{He1_5876_2011}
\end{figure}
\begin{figure}
\includegraphics[width=88mm, trim = 35 0 0 0, clip = true, angle=-90]{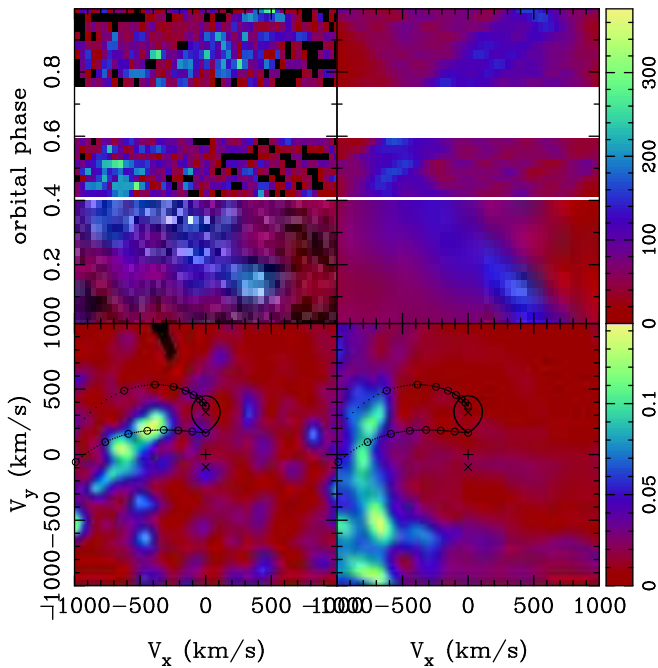}
\caption{Modulation tomogram of the emission line O VI ($\lambda$5290) using 2010 data. Panels are as described in Fig. 9.}
\label{O6_2010}
\end{figure}
\begin{figure}
\includegraphics[width=88mm, trim = 35 0 0 0, clip = true, angle=-90]{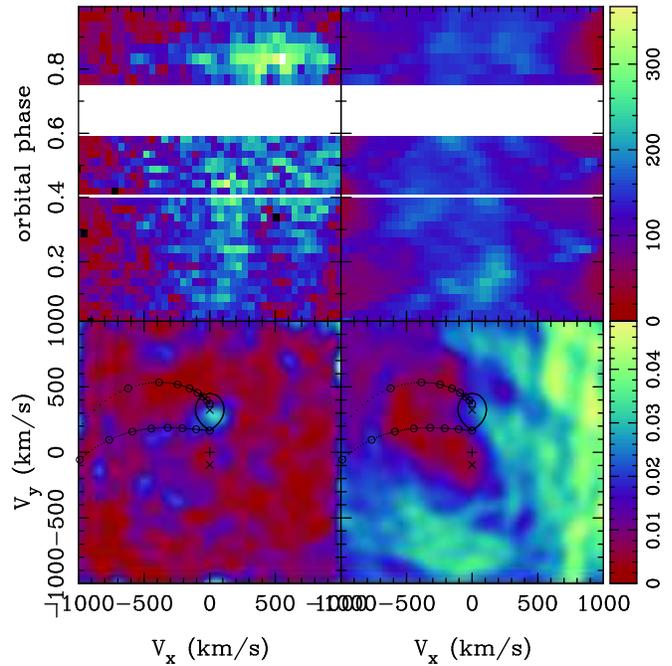}
\caption{Modulation tomogram of the Bowen Blend using 2010 data. Panels are as described in Fig. 9.}
\label{4640_2010}
\end{figure}

\section{Discussion}

\begin{figure}
\includegraphics[width=84mm]{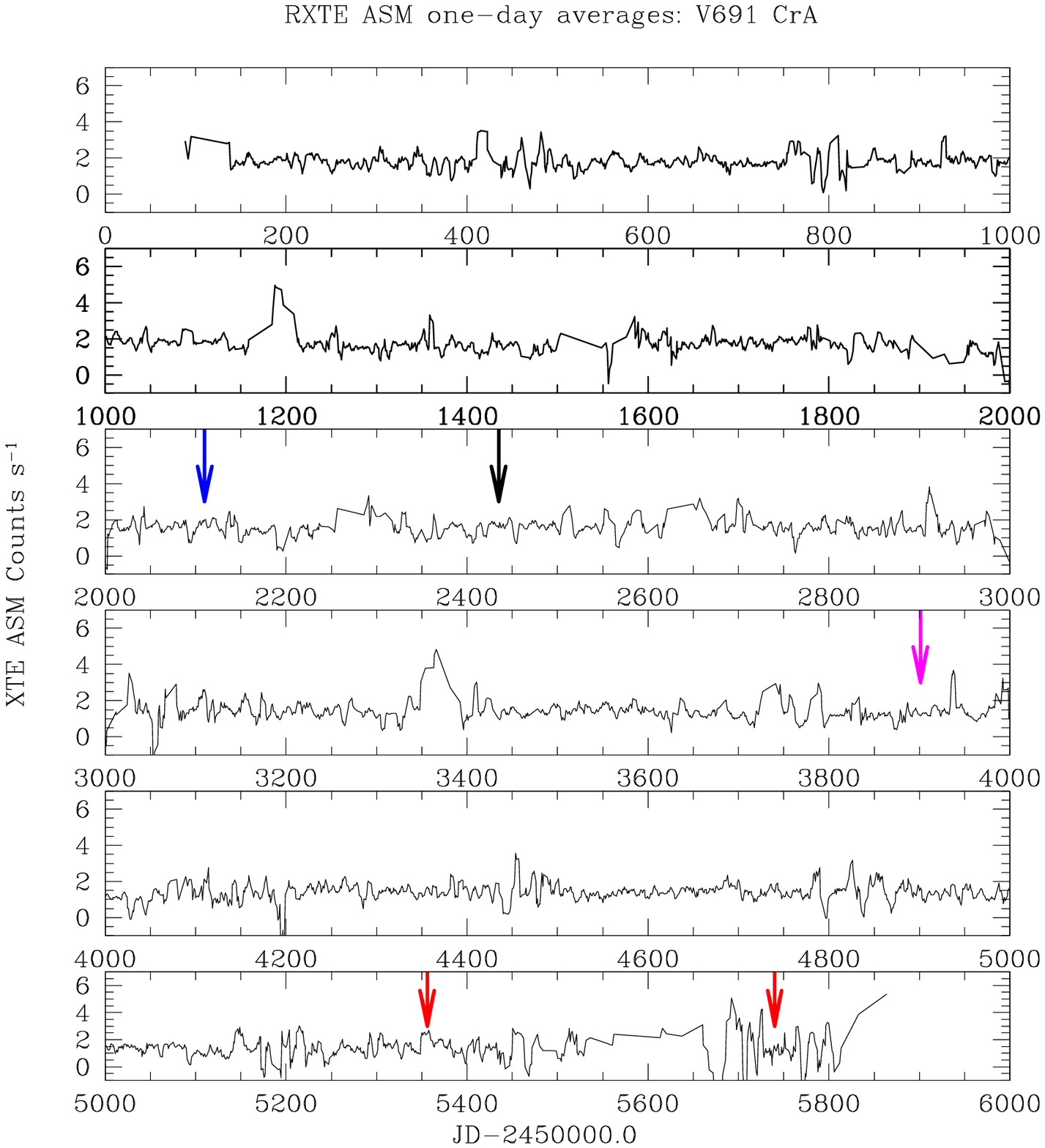}
\caption{The RXTE ASM light curve data for V691 CrA. The blue arrow indicates the time of optical observations taken by \citet{b16}, black indicates the time of optical observations taken by \citet{b6}, magenta indicates the time of optical observations taken by \citet{b30} and the red arrows indicate the times of optical observations used in this paper. This data was retrieved using the ASM data website (courtesy of the MIT ASM team) at $http://xte.mit.edu/ASM\_lc.html$.}
\label{lightcurve}
\end{figure}

He II ($\lambda$5411) displays the highest inner disc velocities while H$\alpha$ displays the lowest inner disc velocities. The disc in He II ($\lambda$5411) seems to be centred on the predicted position of the neutron star while the discs in He II ($\lambda$4686) and H$\alpha$ are offset towards the centre of mass of the system indicating possible eccentricities. The observations imply that He II ($\lambda$5411) emission originates from material closer to the inner disc while H$\alpha$ and He II ($\lambda$4686) emission originate from the material further away from the centre of the disc. 

Our observations of He II ($\lambda$4686 and $\lambda$5411) are consistent with those made by \citet{b30} who observed no eclipsing of the He II flux. They suggest a disc-wind model in which the emission in He II might originate from matter above the accretion disc that still retains some of the velocity field of the disc.

Together with those produced by \citet{b6} and \citet{b30} our tomograms cover the He II ($\lambda$4686) line at four different times as shown in  Fig. 18. We use this chronological data to show that the tomograms exhibit consistent disc wind velocities which span approximately $\pm$500 km s$^{-1}$. Our values of K$_1$ (94.5 km s$^{-1}$) and K$_2$ (324 km s$^{-1}$; value obtained by \citet{b6} for a neutron star mass of 1.4 M$_{\sun}$) give a mass ratio of 0.29. From \citet{b47} (figure 3) we find that for a mass ratio of 0.29, the minimum allowed velocity for a Keplerian disc is approximately 1.8 $\times$ 324 km s$^{-1}$  or 583 km s$^{-1}$. This implies that the disc velocities we observe are sub-Keplerian as suggested by \citet{b30}.

We also find that the structure in the emission ring seen by \citet{b30} matches the structure we detect in our 2010 observations, and that the emission ring structure seen by \citet{b6} matches the structure we observe in 2011 suggesting periodicity in the system. SPH simulations by \citet{b46} show that the manifestations of eccentric discs are bright emission between the gas stream trajectory and the Keplerian velocity of the disc along the gas stream, non-sinusoidal S-waves, crescent-shaped features in Doppler maps, and the shifting of the map centre of symmetry away from (0,-K$_1$). We observe crescent shaped bright regions in our He II tomograms. The ring structure seen in H$\alpha$ and He II ($\lambda$4686) is off-centred from the predicted position of the neutron star. We also note that H$\alpha$ which originates from material close to the disc edge displays an asymmetric structure.  These observations may suggest disc precession in the system which according to \citet{b44}, is to be expected in LMXBs with a mass ratio $\leq$ 0.3 as the disc reaches the 3:1 resonance with the binary companion. However, \citet{b45} did not find any periodicity in this object between 1-100 days. More observations covering regular yet well-spaced time intervals are required in order to resolve this issue.

\subsection{Balmer line absorption and emission}

The deep troughs on either side of H$\alpha$ and H$\beta$ (as seen in Fig. 1) confirm the presence of Balmer line absorbing material in the system. 

Our observations of H$\beta$ are consistent with those made by \citet{b30} and are in agreement with their ‘spray’ model: the narrow blueshifted H$\beta$ absorption feature at phases 0.6-0.7 is interpreted as spraying matter from the hotspot in the direction of the observer, and the smaller redshifted absorption feature after orbital phase 0.1 is interpreted as a spray over the disc. In H$\alpha$ we observe strong absorption over a much larger phase range than in H$\beta$. We also observe that the weakest absorption in H$\alpha$ is seen at phases 0.7-0.8 where a strong absorption dip is present in H$\beta$ (Fig. 2).

Formation of the spray close to the hot spot means the spray is partially shielded from view by the disc bulge \citep{b38} causing only the cooler spray above the bulge to be visible which leads to a sharp dip in H$\beta$ while no dip is present in H$\alpha$. The reason for absorption in a larger phase range in H$\alpha$ than in H$\beta$ could be that the spray traveling downstream will heat up gradually, enabling H$\alpha$ absorption to still occur when H$\beta$ absorption is no longer possible.

The bright spots on the positive V$_y$ axis seen in the Doppler tomograms (Fig. 3 and Fig. 4) and the super-imposed S-curves seen in the modulation tomograms (Fig. 9 and Fig. 10) suggest H$\alpha$ emission from the secondary in addition to the emission from the disc. The amplitude variation map in 2010 shows modulation arising from the bright spot, usually a feature expected from the secondary. The reason for the emission is uncertain. These observations are in agreement with the narrow H$\alpha$ peaks observed by \citet{b39} which they suggested originate from the donor star. H$\alpha$ emission from the companion star has been observed in other systems as well. \citet{b7} observed H$\alpha$ emission from the leading side of the inner face of the companion star in Cen X-4 and attributed it to irradiation from the hotspot and shielding from the gas stream. \citet{b31} observed H$\alpha$ emission from the secondary of J100658.40+233724.4. \citet{b11} detected H$\alpha$ emission from the secondary of the black hole system A0620-00 which they attributed to chromospheric activity induced by rapid rotation.

\subsection{Absorption in He I ($\lambda$5876)}

We detect a bright spot in He I ($\lambda$5876) in absorption which seems to coincide with the L$_1$ point in 2010 while shifting to the leading face of the secondary in 2011. We detect strong bright spots in the same region of the amplitude variation maps indicating sinusoidal variation of the line flux. We also observe that the S-curves in the He I modulation tomograms are opposite in phase to those in the He II tomograms. These observations suggest the point of origin of the He I absorption line to be close to the leading side of the inner face of the secondary. The reason for the shift in the position of the feature in velocity space is uncertain.

Our tomograms show the absorption feature in He I at $\sim$ 189 km s$^{-1}$ on the V$_y$ axis in 2010 and at $\sim$ 279 km s$^{-1}$ in 2011. Since these values obtained using He I are lower than the K$_2$ value obtained by \citet{b6} using the N III line of the Bowen blend, we interpret the bright spots as He I absorption originating near the L$_1$ point from gas leaving the surface of the donor.

\section{Conclusions}

In this paper we present Doppler and modulation tomography of V691 CrA using data taken in 2010 and 2011.

Our observations cover a greater range of wavelength ($\lambda$4400 - $\lambda$6650) than any observations previously used for tomography. This extended range enabled us to produce the first modulation tomograms of He I ($\lambda$5876) and H$\alpha$ ($\lambda$6562.76). We detect the secondary in emission with H$\alpha$. We also detect absorption in He I ($\lambda$5876) close to the L$_1$ point of the secondary.

Our observations in He I are consistent with the observations made by \citet{b13}, \citet{b16}, \citet{b39} and with the interpretation of He I given by \citet{b6}. \citet{b8} observed that, for several dwarf novae with significant eccentricities, the irradiated side of the secondary does not seem to be centred towards the primary but instead towards the leading side of the star. This was especially true for the dwarf novae U Gem and IP Peg. The same form of leading face heating was observed in the AM Her system V824 Cen by \citet{b25}. This phenomena seems to be the case for V691 CrA in 2011.

Our observations of H$\alpha$ and He I ($\lambda$5876) suggest that these lines originate from different regions of the companion star. He I originates near the L$_1$ point from gas leaving the surface of the donor star, N III originates in the lower latitude most directly illuminated regions of the donor star and H$\alpha$ originates at higher latitudes where the donor is illuminated at a steeper angle of incidence (close to the poles).

The deep troughs on either side of H$\alpha$ and H$\beta$ confirm the presence of absorbing material in the system. Our observations of H$\alpha$ show absorption in a larger phase range than in H$\beta$. This could be due to heating up of the spray \citep{b30} as it travels downstream enabling H$\alpha$ absorption to still occur when H$\beta$ absorption is no longer possible. We also suggest possible occultation of the H$\alpha$ absorbing spray by the disc bulge at certain phases.

Where tomograms exist for a given line our 2010 observations are consistent with \citet{b30} and our 2011 observations with \citet{b6}. This suggests repeating variability in the system and shows the need for further observations to see if the variability is periodic.

\section{Acknowledgments}

This research was supported in part by a Smithsonian Endowment Grant awarded to SDV. We would like to thank Tom Marsh for the use of his MOLLY and DOPPLER software packages and Danny Steeghs for the use of his MODMAP software package. CSP would like to thank Tilak and Deepika Peris and Ayesha Casie Chetty for their gracious support and constructive comments. We thank an anonymous referee for constructive comments which greatly improved the quality of the paper.

\appendix

\bsp
\label{lastpage}

\end{document}